\documentclass[%
onecolumn,
nofootinbib,
 amsmath,amssymb,
 aps,
]{revtex4-2}

\usepackage{graphicx}
\usepackage{dcolumn}
\usepackage{bm}
\usepackage{hyperref}

\usepackage{color}

\usepackage[T1]{fontenc}
\usepackage[utf8]{inputenc}
\usepackage{graphicx}
\usepackage[dvipsnames]{xcolor}
\usepackage{hyperref}
\usepackage{mathtools}
\usepackage{booktabs}
\usepackage{array}
\usepackage{tikz}
\usetikzlibrary{arrows.meta,decorations.pathmorphing,decorations.markings,
                positioning,calc,shapes}
\usepackage{caption}
\usepackage{subcaption}

\hypersetup{colorlinks=true,citecolor=blue,linkcolor=blue,urlcolor=blue}

\newcommand{\PV}{\mathrm{PV}}

\newcommand{\dIHO}{\mathrm{dIHO}}
\newcommand{\IHO}{\mathrm{IHO}}
\newcommand{\diag}{\mathrm{diag}}
\newcommand{\mink}{\eta_{\mu\nu}=\diag(-,+,+,+)}

\newcommand{\LF}{\left(}
\newcommand{\RF}{\right)}

\tikzset{
  pvline/.style={thick, decorate,
    decoration={snake, amplitude=1.8pt, segment length=5pt, post length=2pt}},
    pvline1/.style={thick, decorate,
    decoration={snake, amplitude=2pt, segment length=8pt, post length=1pt}},
  feynline/.style={thick},
  cutline/.style={dashed, thick, red!80!black},
  vtx/.style={circle, fill=black, inner sep=1.5pt},
}

\begin{document}

\title{Unitary Quadratic Quantum Gravity in 4D}

\author{K. Sravan Kumar}
\email{sravan.kumar@port.ac.uk}
\affiliation{Institute of Cosmology \& Gravitation,
	University of Portsmouth,
	Dennis Sciama Building, Burnaby Road,
	Portsmouth, PO1 3FX, United Kingdom}

\author{Jo\~ao Marto}
\email{jmarto@ubi.pt}
\affiliation{ Departamento de F\'isica, Centro de Matem\'atica e Aplicações (CMA-UBI), Universidade da Beira Interior, Rua Marquês D'Ávila e Bolama, 6201-001 Covilhã, Portugal}

\date{\today}

\begin{abstract}
In quadratic gravity, with a positive Weyl squared coefficient, the extra spin–2 sector is shown to correspond to a dual inverted harmonic oscillator, instead of a ghost. Using the Wightman spectrum condition, we prove that the associated K\"{a}ll\'{e}n--Lehmann spectral density vanishes, reflecting the absence of a normalizable ground state and the spacelike nature of the propagator pole. This uniquely fixes the propagator to a principal value form as a theorem, not a prescription. The optical theorem is satisfied, the dual IHO spin-2 is not an asymptotic state, and gives only virtual contributions at all loop orders. As a result, unitarity is preserved consistently with renormalizability.
\end{abstract}

\maketitle

\section{Introduction}
\label{sec:intro}
Quadratic (Stelle) gravity~\cite{Stelle1977},
\begin{equation}
	S=\int\! d^4x\sqrt{-g}\!\left[\frac{M_p^2}{2}R+\frac{\alpha}{2}R^2
	+\frac{\beta}{2}W_{\mu\nu\rho\sigma}W^{\mu\nu\rho\sigma}\right],
	\label{eq:action}
\end{equation}
is the unique perturbatively renormalizable local theory of quantum gravity in four
spacetime dimensions~\cite{Stelle1977,Salvio2018,Buoninfante2025}. This uniqueness is not an
accidental technicality. In quantum field theory, perturbative renormalizability is the criterion
that isolates predictive ultraviolet (UV) completions from merely effective low-energy descriptions,
which is precisely why it underlies the success of the Standard Model~\cite{WeinbergQFT1}.
For gravity, the higher derivative terms in~\eqref{eq:action} raise the UV falloff of the
graviton propagator from $1/k^2$ to $1/k^4$, thereby yielding power counting renormalizability
in four dimensions~\cite{Stelle1977,Salvio2018}. The standard obstruction, however, is that the
additional spin-2 degree of freedom has long been interpreted as a ghost, apparently spoiling
unitarity~\cite{Stelle1977,Salvio2018}. A variety of proposals have been developed to evade this
difficulty, including fakeons~\cite{Anselmi2018}, Lee--Wick-type prescriptions and related contour
deformations~\cite{LeeWick,Donoghue2019}, and PT-symmetric quantization~\cite{Bender2008},
all of which modify the interpretation of the extra pole while introducing structure beyond the
minimal local theory.

In our companion work~\cite{KumarMarto2026}, we argued that this standard conclusion is not
forced upon quadratic gravity. For the sign choice $\beta>0$, the extra spin-2 sector is not
governed by a negative definite ghost Hamiltonian, but by a dual inverted harmonic
oscillator (dual-IHO), namely an indefinite Hamiltonian system with hyperbolic phase
space which shares close similarities with IHO. This distinction is decisive. A ghost is pathological because its negative definite Hamiltonian leads to negative norm excitations and a direct conflict with the probabilistic interpretation of quantum theory. By contrast, the IHO and dual-IHO are not negative definite systems (but indefinite); their essential quantum property is instead the absence of a normalizable ground
state. The physical relevance of this structure is not peculiar to quadratic gravity. IHO dynamics is a universal template for controlled instability: it appears in the
Berry--Keating $H=xp$ system~\cite{BerryKeating1999}, and in our broader framework it also
underlies the quantization of the quadratic gravity spin-2 sector~\cite{KumarMarto2026}. The point of the present paper is therefore not to introduce an ad hoc prescription, but to pinpoint the precise mechanism by which this universal IHO structure restores unitarity in quadratic gravity. The key issue is the propagator of the extra spin-2 field. For an ordinary particle degree of
freedom, the full propagator is constrained by the K\"all\'en--Lehmann (KL)
representation~\cite{Kallen1952,Lehmann1954}, whose spectral density is defined on the timelike
spectrum of physical states and is rooted in the Wightman spectrum condition of local quantum
field theory~\cite{Streater1964,Haag1996}. In the dual-IHO case, however, two facts hold
simultaneously: the field admits no normalizable vacuum state, and its pole lies at spacelike
momentum. These two facts are already sufficient to force the KL spectral density to vanish.
Consequently, the propagator of the dual-IHO spin-2 field is fixed uniquely to its
principal value form (PV). In other words, the principal value is not an externally imposed prescription; it is a theorem dictated by the Hilbert space structure and the spectral support of
the theory~\cite{KumarMarto2026}.

In this paper, we sharpen that conclusion into a self-contained unitarity analysis with three main
results. First, we show that the KL spectral density vanishes as a consequence of the Wightman
spectrum condition applied to a field with spacelike momentum support, so that $\rho=0$ is not
an assumption but a theorem. Second, we give a Cutkosky analysis about how the optical theorem is satisfied in our case, with an additional spin-2 field not corresponding to any on-shell states. Third, we perform the Landau analysis of the
relevant bubble diagrams and compare it explicitly with the Feynman--Wheeler case of
Ref.~\cite{Anselmi2020}, thereby showing why the nonlocal divergences associated with timelike
principal value propagators are absent when the pole is spacelike. The result is a minimal and
sharp statement: in quadratic gravity with $\beta>0$, the extra spin-2 mode remains purely
virtual, contributes to UV renormalization through its dispersive part, but never enters
physical absorptive cuts. Unitarity and renormalizability, therefore, coexist within the original
local theory~\cite{KumarMarto2026}.

We use the metric signature $\mink$, so that $k^2=-(k^0)^2+|\vec{k}|^2$. Timelike momenta
satisfy $k^2<0$, while spacelike momenta satisfy $k^2>0$. We work in natural units $\hbar=c=1$ and $8\pi G = M_p^{-2}$.
\section{The dual-IHO: Berry--Keating and phase space structure}
\label{sec:dIHO}
\begin{figure}[t]
\centering
\includegraphics[width=0.80\columnwidth]{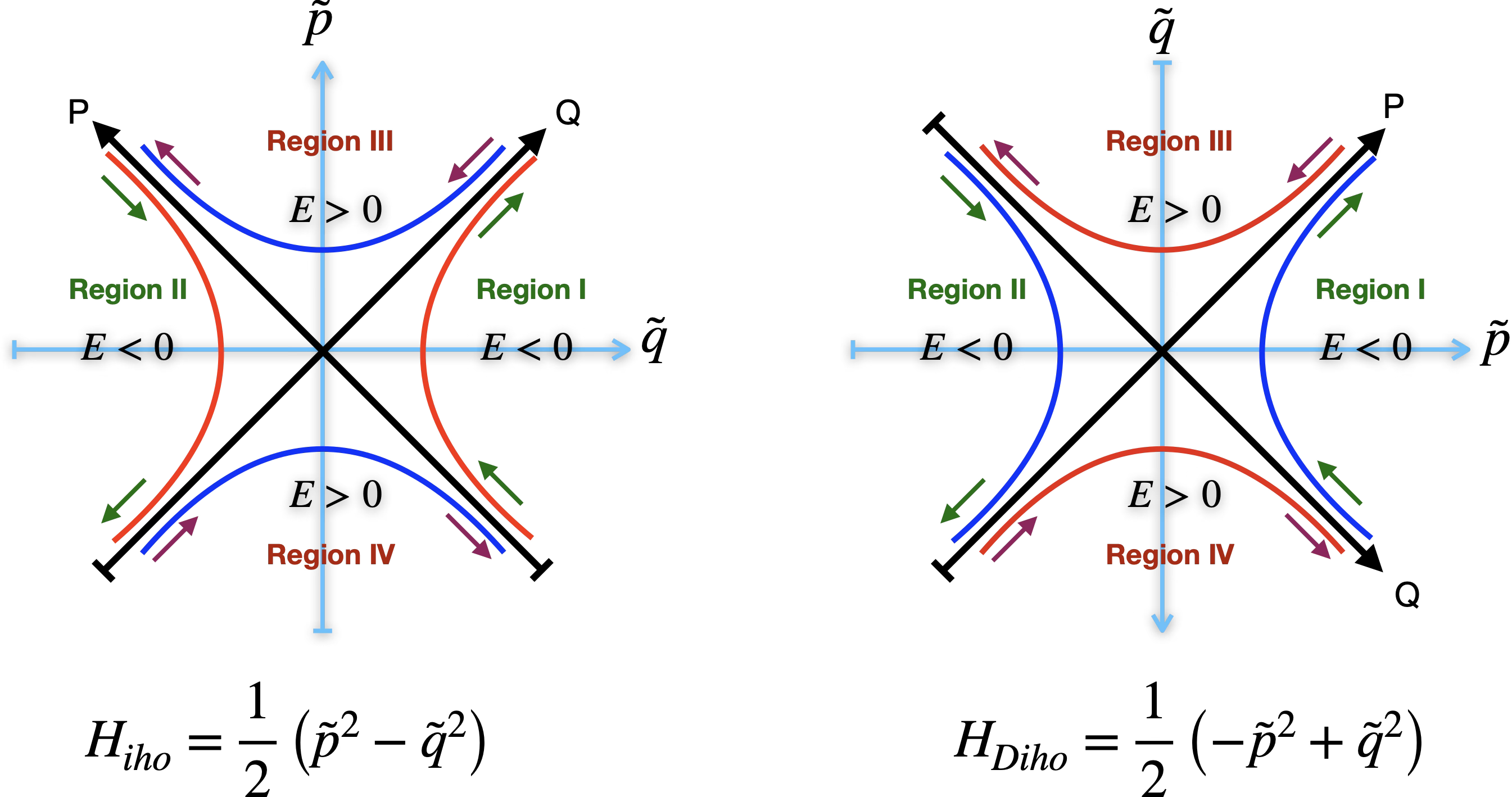}
\caption{Phase space of the IHO (left, $H_{\IHO}=\frac\omega2(\tilde p^2-\tilde q^2)$)
and the dual-IHO (right, $H_{\dIHO}=\frac\omega2(-\tilde p^2+\tilde q^2)$). Both are related by a 90-degree canonical rotation and share hyperbolic phase-space trajectories.
The separatrices divide each phase space into four regions with two distinct arrows of
time, requiring a direct-sum quantization framework~\cite{KumarMarto2026,Kumar2025dS,
Kumar2024BH}. Neither system is a ghost.}
\label{fig:phasespace}
\end{figure}

The full graviton propagator of \eqref{eq:action} around Minkowski spacetime, 
in de~Donder gauge, can be written with the spin projectors $\LF P^{(2)}_{\mu\nu,\rho\sigma}\RF$ (spin--2)
and $\LF P^{(0\text{s})}_{\mu\nu,\rho\sigma}\RF $ (spin-0) as \cite{Stelle1977}
\begin{equation}
D_{\mu\nu,\rho\sigma}(k)
=
-\frac{i}{M_p^2}
\left[
\frac{P^{(2)}}{k^2}
-\frac{P^{(2)}}{k^2 -\frac{M_p^2}{\beta}}
-\frac{1}{2}\frac{P^{(0\text{s})}}{k^2}
+\frac{1}{2}\frac{P^{(0\text{s})}}{k^2 + m^2}
\right].
\label{eq:fullprop}
\end{equation}
where
\begin{equation}
 \frac{M_p^2}{\beta} = -m_2^2\,\,\LF \text{if }\beta<0 \RF, \qquad \frac{M_p^2}{\beta} = \mu_2^2 \,\,\LF \text{if } \beta>0 \RF,\qquad
m^2 = \frac{M_p^2}{6\alpha}.
\end{equation}
where $m^2>0$ is mass square of the scalaron, and $m_2^2> 0$ (or $\mu_2^2>0$) is the mass scale associated with the additional spin-2 mode of the theory. The theory has a massless spin-2 and a massive scalar (for $\alpha>0$) (i.e., scalaron) as an on-shell (also off-shell) propagating degree of freedom. The status of additional spin-2 depends on the coefficient of the Weyl square. If $\beta<0$, the spin-2 becomes a pathological ghost and can be an on-shell mode, whereas in the $\beta>0$ case, it becomes a dual IHO spin-2 field, which is a purely off-shell mode as we explain below.
{
In the first case, ($\beta<0$), the extra spin--2 pole is timelike,
\[
k^2=-m_2^2<0,
\]
and this is the pole that is usually identified with the massive spin--2 ghost of quadratic gravity. In the second case, ($\beta>0$), the pole is instead located at
\[
k^2=\mu_2^2>0 .
\]
Thus the two sign choices do not merely differ by the sign of a residue. They place the additional spin--2 pole on different Lorentz-invariant momentum shells. The ($\beta<0$) pole lies in the usual timelike spectral domain, whereas the ($\beta>0$) pole lies on a spacelike shell. This invariant distinction is the starting point of the analysis below, where we examine whether the ($\beta>0$) sector can be assigned a physical particle interpretation.}The four systems relevant to quadratic gravity are distinguished by their Hamiltonians:
\begin{align}
H_{\rm HO}&=\tfrac\omega2(\tilde p^2+\tilde q^2),
\quad\text{positive definite},\label{eq:HO}\\
H_{\rm ghost}&=-\tfrac\omega2(\tilde p^2+\tilde q^2),
\quad\text{negative definite},\label{eq:ghost}\\
H_{\IHO}&=\tfrac\omega2(\tilde p^2-\tilde q^2),
\quad\text{indefinite},\label{eq:IHO}\\
H_{\dIHO}&=\tfrac\omega2(-\tilde p^2+\tilde q^2),
\quad\text{indefinite}.\label{eq:dIHO}
\end{align}
The IHO and dual-IHO Hamiltonians are indefinite, not negative definite.
This single fact separates them categorically from the ghost. In the canonically rotated
coordinates $Q=(\tilde p+\tilde q)/\sqrt{2}$, $P=(\tilde p-\tilde q)/\sqrt{2}$, both
take the Berry--Keating (BK) form~\cite{BerryKeating1999}:
\begin{equation}
H_{\IHO}=\tfrac\omega2(QP+PQ),\qquad
H_{\dIHO}=-\tfrac\omega2(QP+PQ).
\label{eq:BK}
\end{equation}
The BK Hamiltonian $H=xp$ is famous for its connection to the Riemann
hypothesis~\cite{BerryKeating1999}: its energy spectrum, under appropriate quantization
conditions, reproduces the non-trivial zeros of the Riemann zeta function. Both the IHO
and dual-IHO share hyperbolic classical trajectories $Q=Q_0e^{\omega t}$,
$P=P_0e^{-\omega t}$, and the same phase-space structure up to a 90-degree canonical
rotation (Fig.~\ref{fig:phasespace}). The dual-IHO phase space is an equivalent physical
system with the roles of positive and negative energy regions swapped.
The quantization of the IHO/dual-IHO poses a fundamental challenge, namely the fact that phase-space regions with opposite arrows of time (separated by the separatrices in
Fig.~\ref{fig:phasespace}) cannot be consistently handled by a single Schr\"{o}dinger equation. This is precisely the open problem identified by Berry and Keating~\cite{BerryKeating1999}.
The resolution is provided by direct-sum quantum field theory
(DQFT)~\cite{KumarMarto2026,Kumar2025dS,Kumar2024BH,Kumar2024Rindler,Kumar2024Hawking,
Gaztanaga2026}, which quantizes the IHO on a split Hilbert space
$\mathcal H=\mathcal H_+\oplus\mathcal H_-$ with geometric superselection sectors
corresponding to parity conjugate regions. The BK antipodal identification $(Q,P)\sim(-Q,-P)$
then acquires a geometric interpretation via the superselection structure. In this paper,
we use only the consequence that the dual-IHO admits no global Fock vacuum.
For quadratic gravity, with $\beta>0$, the transverse and traceless (TT) spin-2 sector after diagonalization gives
the Hamiltonian~\cite{KumarMarto2026}
\begin{equation}
H_{\rm TT}=\frac12\sum_{s}\int\!\frac{d^3k}{(2\pi)^3}
\bigl(\dot u_s^2+\textbf{k}^2u_s^2\bigr)
-\frac12\sum_{s}\int\!\frac{d^3k}{(2\pi)^3}
\bigl(\dot v_s^2+(\textbf{k}^2-\mu_2^2)v_s^2\bigr),
\label{eq:HTT}
\end{equation}
where $\mu_2^2\equiv M_P^2/\beta>0$. The field $u_s$ is the massless graviton and each
mode of $v_s$ takes the form 
\begin{equation}
H_{s,\lambda}=-\tfrac12 p_{s,\lambda}^2+\tfrac12(-\lambda+\mu_2^2)q_{s,\lambda}^2,
\label{eq:mode}
\end{equation}
where 
\begin{equation}
v_s(t,\vec x)
=
\sum_\lambda q_{s,\lambda}(t) f_\lambda(\vec x),
\qquad
\pi_s(t,\vec x) = -\dot v_s
=
-\sum_\lambda p_{s,\lambda}(t) f_\lambda(\vec x).
\end{equation}
with 
\begin{equation}
-\nabla^2 f_\lambda(\vec x)
=
\lambda\, f_\lambda(\vec x),
\qquad
\lambda \ge 0,
\end{equation}
Notice that the Hamiltonian \eqref{eq:mode} is exactly a dual-IHO system. Moreover, $v_s$ is a dual-IHO field, which can be understood as an infinite collection of dual-IHOs, with no global Fock vacuum.
 
{A potential concern is that the mode Hamiltonian corresponding to \eqref{eq:mode} appears to admit a momentum space splitting into regions with qualitatively different behavior. One may therefore wonder whether a subset of modes can be quantized as ordinary harmonic oscillators with normalizable ground states, thereby restoring a conventional particle interpretation. We now show that such a splitting is not Lorentz invariant and therefore cannot define a physical decomposition of the theory. 
We caution that one cannot split the modes based on the frame-dependent $\textbf{k}^2$ terms, i.e., one may be tempted to divide the momentum space of the dual-IHO hamiltonian into regions
\begin{equation}
    \textbf{k}^2 < \mu_2^2,\quad   \textbf{k}^2 > \mu_2^2
    \label{splitmodes}
\end{equation}
and interpret the Hamiltonian densities as indefinite or negative definite. This would enable to define a normalizable ground state for the dual IHO field and adopt the KL representation. Such a decomposition cannot define distinct physical sectors because the inequalities \eqref{splitmodes} are not Lorentz invariant. Physical sectors in a relativistic theory must be characterized by Lorentz invariant quantities\footnote{{To be more precise,  let us apply Lorentz boost transformations in the \textbf{x}-direction 
\begin{equation}
\begin{split}
k'^0 &=\gamma\,(k^0-v\,k^1)\\
k'^1 &=\gamma\,(-vk^0+\,k^1), \qquad \gamma=(1-v^2)^{-1/2}.  
\label{eq:boost}
\end{split}
\end{equation}
by taking $k^\mu = (k^0, k^1,0,0)$. Since only $k_{\nu}k^{\nu}$ is Lorentz invariant, a mode satisfying $(k^1){^2}>\mu_2^2$  in one frame can satisfy $(k^{\prime 1}){^2}<\mu_2^2$  in another, for a suitable choice of $v<1$. The Lorentz group, therefore, mixes the two regions. Neither region is separately invariant, and neither can be assigned an independent physical interpretation. Therefore, the dual-IHO sector cannot be viewed as a mixture of ordinary oscillators (i.e., ghosts with a negative definite Hamiltonian) and inverted oscillators classified by $\textbf{k}^2 \lessgtr \mu_2^2$. Such a classification is frame-dependent and is not preserved by Lorentz transformations. The physically relevant object is the full Lorentz invariant orbit associated with the spacelike pole, not a frame-dependent partition of its Fourier modes. A normalizable vacuum constructed from the subset $\textbf{k}^2 > \mu_2^2$  would itself fail to be Lorentz invariant because Lorentz boosts map modes into and out of that subset. The construction of \cite{Jacobson:1987ap} relies on selecting by hand the subset $\textbf{k}^2>\mu_2^2$ in order to define a normalizable vacuum. Since this selection is not Lorentz invariant, that framework is conceptually different from the present investigation, where Lorentz covariance is maintained throughout.}}, such as the orbit determined by the invariant four momentum $k^2$, rather than by the frame dependent three momentum magnitude $|\textbf{k}|$ (See \cite{KumarMarto2026} for more detailed discussion).}

\section{K\"{a}ll\'{e}n--Lehmann representation with a vanishing spectral density}
\label{sec:KL}

The full nonperturbative Feynman propagator of any scalar field admits the K\"{a}ll\'{e}n--Lehmann (KL) representation~\cite{Kallen1952,Lehmann1954}
\begin{equation}
\widetilde\Delta(k^2)=\frac{-i}{k^2+m^2-i\epsilon}
+\int_0^\infty\!ds\,\frac{-i\rho(s)}{k^2+s-i\epsilon},
\label{eq:KL}
\end{equation}
with spectral density $\rho(s)=\sum_n|\langle k,n|\phi(0)|0\rangle|^2\delta(s-M_n^2)\geq0$.
The lower limit $s=M_n^2\geq0$ is not an independent assumption. It is a direct consequence
of the Wightman spectrum condition ~\cite{Streater1964,Haag1996}:
\begin{equation}
\mathrm{spec}(\hat P^\mu)\subset\bar V^+=
\{k^\mu:\,k^2\leq0,\;k^0\geq0\},
\label{eq:spectrum}
\end{equation}
which requires every physical state $|n\rangle$ to satisfy $M_n^2=-k_n^2\geq0$.
The IHO and dual IHO fields with spacelike poles are excluded from the KL domain a priori
by the spectrum condition. Due to this reason, the studies of QFT with tachyons in \cite{Jacobson1988,Jodlowski:2026zaw} are not applicable in our consideration.   
Therefore, since the spectral parameter $s$ is restricted to non-negative values, the vanishing of the spectral density for the dual-IHO spin--2 field, $\rho(s)=0$, follows from two independent reasons. 

\textbf{Reason 1 (absence of a normalizable vacuum):} The KL matrix element
$\langle0|\phi(0)|n\rangle$ requires a normalizable bra $\langle0|$ with $\langle0|0\rangle=1$.
The eigenvalue equation of the dual-IHO,
\begin{equation}
\tfrac\omega2(\tilde q^2+\partial_{\tilde q}^2)\psi_E=E\psi_E,
\label{eq:IHO_eq}
\end{equation}
has solutions that are parabolic cylinder functions $D_\nu(\tilde q)$ with
$\nu=iE/\omega-\frac12$. Their asymptotics~\cite{KumarMarto2026}
\begin{equation}
D_\nu(\tilde q)\sim \tilde q^\nu e^{-\tilde q^2/4}
+\frac{\Gamma(-\nu)}{\sqrt{2\pi}}\tilde q^{-\nu-1}e^{+\tilde q^2/4},
\quad\tilde q\to-\infty,
\label{eq:asymp}
\end{equation}
show that the second branch grows as $e^{+\tilde q^2/4}$: not square-integrable for
any real energy $E$. No eigenstate of the dual-IHO Hamiltonian belongs to
$L^2(\mathbb R)$; no normalizable vacuum exists. The KL matrix element is undefined,
and one must set $\rho(s)=0$.

\textbf{Reason 2 (spacelike pole):} The field equation of the $v_s$ sector,
$\ddot v_s-\nabla^2v_s-\mu_2^2v_s=0$, gives a propagator pole at
\begin{equation}
k^2=\mu_2^2>0\quad\text{(spacelike in $-{+}{+}{+}$)}.
\label{eq:pole}
\end{equation}
In the KL integral \eqref{eq:KL}, this would require $s=-\mu_2^2<0$, outside the spectral
domain $s\geq0$. It would correspond to a physical state with imaginary invariant mass, in direct violation of the spectrum condition \eqref{eq:spectrum}, a situation which is inadmissible. Therefore, both reasons independently enforce $\rho=0$. 
The $\delta(k^2-\mu_2^2)$ piece is absent identically from the propagator, not suppressed by kinematics, not removed by prescription, but absent as a theorem. The propagator retains only its dispersive part,
\begin{equation}
\Delta_\dIHO(k)=\PV\!\left(\frac{i}{k^2-\mu_2^2}\right).
\label{eq:PV}
\end{equation}
This is the unique propagator compatible with the KL representation and the Hilbert-space structure of the dual-IHO sector. It is not a fakeon projection, not a Lee--Wick contour, and not a Feynman--Wheeler prescription {(See Appendix~\ref{subsec:not_FW_prescription} for more details)}.
Let us notice that both the Feynman propagator $F=-i/(k^2-\mu_2^2-i\epsilon)$ and the anti-Feynman propagator $\bar F=i/(k^2-\mu_2^2+i\epsilon)$ require $\rho\neq0$ to be defined in the KL representation. Here, in the context of a dual IHO spin-2 field, $\rho=0$ is a consequence of the lack of a normalized ground state and the spacelike nature of the pole. One 
cannot define a vacuum for the dual IHO spin-2 field, this is completely an off-shell degree of freedom participating in the virtual interactions of the massless gravitation field of the quadratic gravity. 
Therefore, the identity $\PV=\frac12(F+\bar F)$ (a distributional identity which is algebraically valid) is physically not possible to define for the dual-IHO spin-2 field. This is the precise reason why the objections of~\cite{Anselmi2020,Jacobson1988} about timelike momentum supported PV propagators do not apply to the present case. We note that recent work on tachyons \cite{Jodlowski:2026zaw} with PV propagator ignores the subtleties with the KL representation associated with IHO and dual IHO fields. 
Table~\ref{tab:comparison} summarizes the key distinctions.

\begin{table}[t]
\centering
\renewcommand{\arraystretch}{1.35}
\begin{tabular}{lcccc}
\toprule
System & Hamiltonian & Ground state & Pole & $\quad \rho$\\
\midrule
HO      & pos.\ def.  & $L^2$ yes & timelike $k^2<0$ & $\quad \neq0$\\
Ghost   & neg.\ def.  & $L^2$ yes & timelike $k^2<0$ & $\quad \neq0$\\
IHO     & indefinite  & $L^2$ no  & spacelike $k^2>0$ & $\quad=0$\\
dual-IHO& indefinite  & $L^2$ no  & spacelike $k^2>0$ & $\quad=0$\\
\bottomrule
\end{tabular}
\caption{Key properties in $-{+}{+}{+}$ signature. The IHO and dual-IHO share
spacelike poles and no normalizable ground state; both independently force $\rho=0$.}
\label{tab:comparison}
\end{table}


To be more specific, if we apply the Cutkosky cutting rule to the scattering process in quadratic gravity $\beta>0$, the optical theorem $2\,\mathrm{Im}\,\mathcal M(a\to a)=\sum_f\int d\Pi_f|\mathcal M(a\to f)|^2$ is trivially satisfied. 
Since the spin-2 IHO field propagator is \eqref{eq:PV}, as its propagation is for space-like momenta, as such forbidden to be an on-shell state in any kinematic scattering process, and since its spectral density also vanishes, it contributes zero imaginary part to any amplitude. For the ghost ($\beta<0$, timelike pole at $k^2=-m_2^2<0$), the tree-level amplitude for $2\to2$ graviton scattering gives
\begin{equation}
2\,\mathrm{Im}\,\mathcal M_{\rm ghost}
=-\frac{2\pi \mathcal{N}}{M_P^2}\,\delta(k^2+m_2^2)<0,\quad \mathcal{N}>0,
\label{eq:ghost_violation}
\end{equation}
where $\mathcal{N} = V^{\mu\nu\rho\sigma} P^{(2)}_{\mu\nu\alpha\beta} V^{\alpha\beta}_{\rho\sigma} > 0$ collects all real positive kinematic factors from the vertices and spin-2 projector contracted on physical external graviton polarizations. We can clearly see the spin-2 ghost 
is in direct violation of the optical theorem. 
Since the PV propagator \eqref{eq:PV} is the only propagator the dual-IHO spin-2 field admits, every amplitude in which it participates be at the tree level or at any loop order, takes the form
\begin{equation} 
    \mathcal M_{\rm dIHO}\propto {\rm PV}\big[\frac{1}{k^2-\mu_2^2}\big]
\end{equation}
Hence, no imaginary part and no optical theorem violation. Thus, the Cutkosky rules do not place the additional spin-2 as on-shell states.
In the context of $2\to 2$ forward graviton scattering,  $s=(p_1+p_2)^2<0$ (timelike),
since the dual-IHO pole sits at $k^2=\mu_2^2>0$ (spacelike): the on-shell condition is kinematically inaccessible. Therefore, the optical theorem is saturated only by the massless graviton and the scalaron. Thus, the dual IHO spin-2 field dictates the massless graviton scatterings where it is virtually involved i.e., $t-, u-$ channels, and at all loop levels. 

\section{Full Landau-equation review}
\label{sec:landau}
Having established that the dual-IHO propagator is purely dispersive, we now verify that no physical singularities arise at the loop level. A Landau-equation analysis of the relevant bubble diagrams confirms that the spacelike pole structure prevents any real kinematic pinch, and that the nonlocal UV divergences found by Anselmi \cite{Anselmi2020} for timelike principal value propagators are structurally absent in our case.

\begin{figure}[t]
\centering
\begin{tikzpicture}[scale=1, >=stealth]

\begin{scope}[xshift=0cm]
\draw[pvline1] (-1.1,0) node[left]{\small $P$} -- (-0.0,0);
\draw[pvline1] (3.0,0) -- (4.1,0);
\node[vtx] (L) at (0,0) {};
\node[vtx] (R) at (3,0) {};
\draw[pvline] (L) .. controls +(60:1.8) and +(120:1.8) .. (R);
\draw[pvline] (L) .. controls +(-60:1.8) and +(-120:1.8) .. (R);
\node[above] at (1.5, 1.20) {\footnotesize $\PV(\mu_2)$};
\node[below] at (1.5,-1.20) {\footnotesize $\PV(\mu_2)$};
\node at (1.5,-2.0) {(a)};
\node at (1.5,-2.45) {\footnotesize $P^2_*=4\mu_2^2>0$};
\node at (1.5,-2.85) {\footnotesize (spacelike threshold)};
\end{scope}

\begin{scope}[xshift=8cm]
\draw[pvline1] (-1.1,0) -- (-0.0,0);
\draw[pvline1] (3.0,0) -- (4.1,0);
\node[vtx] (L) at (0,0) {};
\node[vtx] (R) at (3,0) {};
\draw[feynline] (L) .. controls +(60:1.8) and +(120:1.8) .. (R);
\draw[pvline]   (L) .. controls +(-60:1.8) and +(-120:1.8) .. (R);
\node[above] at (1.5, 1.20) {\footnotesize $F(m)$};
\node[below] at (1.5,-1.20) {\footnotesize $\PV(\mu_2)$};
\node at (1.5,-2.0) {(b)};
\node at (1.5,-2.45) {\footnotesize $\alpha_2/\alpha_1 = i\mu_2/m$};
\node at (1.5,-2.85) {\footnotesize (complex: no physical pinch)};
\end{scope}

\end{tikzpicture}
\caption{Bubble diagrams representing 1-loop self energy of a propagating massless spin-2 field (external big wavy lines) with momentum $P$.  The loop wavy lines represent dual-IHO PV propagators. The solid line on the right panel denotes ordinary
Feynman propagator of scalar field (the massive scalaron in our case) (a) Dual-IHO bubble: spacelike branch point $P^2_*=4\mu_2^2>0$.
(b) Mixed bubble: complex Landau parameters, no physical singularity.}
\label{fig:bubbles}
\end{figure}
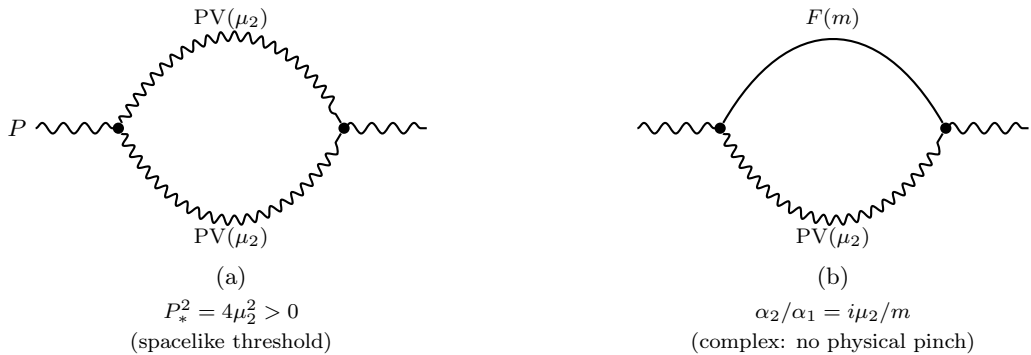

A physical Landau singularity (branch point) requires all Feynman parameters $\alpha_i$ to be simultaneously real and non-negative. We analyze below three distinct bubble configurations, with two involving the dual-IHO, respectively shown in Fig.~\ref{fig:bubbles}.

\begin{enumerate}

\item {\bf Dual-IHO bubble (two spacelike PV lines):} Both PV lines (in Fig.~\ref{fig:bubbles}(a)) have pole at $k^2=\mu_2^2>0$. The Landau on-shell conditions are $k^2=\mu_2^2$ and $(P-k)^2=\mu_2^2$. Subtracting: $P\cdot k=P^2/2$. Writing $k^\mu=aP^\mu+k_\perp^\mu$ with $P\cdot k_\perp=0$ gives $a=1/2$, then
\begin{equation}
k_\perp^2=\mu_2^2-\frac{P^2}{4}.
\label{eq:kperp}
\end{equation}
The branch point (pinch at $k_\perp^\mu=0$) is at
\begin{equation}
P^2_*=4\mu_2^2>0\quad\text{(which is spacelike).}
\label{eq:BP}
\end{equation}
For physical timelike $P^2<0$: $k_\perp^2=\mu_2^2-P^2/4>\mu_2^2>0$ always.
The simultaneous on-shell conditions cannot be satisfied and no absorptive part is generated.

\item {\bf Mixed bubble (one timelike Feynman, one spacelike PV)}. 
Let us consider here a one-loop with a Feynman propagator of the scalaron, with a pole at $k^2=-m^2<0$, and a dual-IHO spin-2 PV with a pole at $k^2=\mu_2^2>0$ (See Fig.~\ref{fig:bubbles}(b)). The Landau equations of this physical process can be written as 
\begin{equation}
\alpha_1(k^2+m^2)=0,\quad
\alpha_2\bigl((P-k)^2-\mu_2^2\bigr)=0,\quad
\alpha_1k^\mu=\alpha_2(P-k)^\mu.
\label{eq:Leq_mixed}
\end{equation}
The momentum equation gives $k^\mu=\frac{\alpha_2}{\alpha_1+\alpha_2}P^\mu$, so
\begin{equation}
\left(\frac{\alpha_1}{\alpha_1+\alpha_2}\right)^{\!2}P^2=-m^2,\quad
\left(\frac{\alpha_2}{\alpha_1+\alpha_2}\right)^{\!2}P^2=\mu_2^2.
\label{eq:Leq_sys}
\end{equation}
Dividing:
\begin{equation}
\frac{\alpha_2^2}{\alpha_1^2}=-\frac{\mu_2^2}{m^2}<0
\;\Longrightarrow\;\frac{\alpha_2}{\alpha_1}=i\frac{\mu_2}{m} \quad(\frac{\mu_2}{m}\in \mathbb R).
\label{eq:complex_L}
\end{equation}
The ratio is purely imaginary. No real non-negative solution exists for any real $P^2$. The threshold lies in the complex $P^2$ plane; therefore, no physical Landau singularity and no nonlocal divergence are present.
    
\item {\bf Feynman--Wheeler mixed bubble (Anselmi~\cite{Anselmi2020}):} Here we consider the mixed bubble diagram (Fig.~\ref{fig:bubbles}(b)) with the notation used in \cite{Anselmi2020} with the context of ghost where both lines have timelike poles at $k^2=-m_i^2<0$ (one Feynman, one with the PV/FW prescription). The same Landau analysis gives
\begin{equation}
\left(\frac{\alpha_i}{\alpha_1+\alpha_2}\right)^{\!2}P^2=-m_i^2<0,\quad i=1,2,
\label{eq:Leq_FW}
\end{equation}
from which
\begin{equation}
\frac{\alpha_2}{\alpha_1}=\frac{m_2}{m_1}>0\quad\text{(real and positive).}
\label{eq:real_L}
\end{equation}
A physical Landau singularity exists at the real timelike threshold $P^2=-(m_1\pm m_2)^2<0$, accessible for physical external momenta. This is the origin of the nonlocal UV divergence found in~\cite{Anselmi2020}
\begin{equation}
\Sigma'_{\rm div}(P)\propto\frac{m_1^2-m_2^2}{P^2}\,\ln\Lambda^2,
\label{eq:nonlocal}
\end{equation}
which cannot be absorbed by a local counterterm and destroys renormalizability. The physical mechanism is clear. Anselmi's nonlocal divergence requires the coexistence of $+i\epsilon$ and $-i\epsilon$ at timelike poles in the same loop. When one pole is spacelike, eq.~\eqref{eq:Leq_sys} gives opposite signs on the right-hand sides, forcing $\alpha_2/\alpha_1$ into the complex plane. No real configuration puts both lines on-shell simultaneously. Table~\ref{tab:landau} summarizes all three cases.
\end{enumerate}

\subsection{Second type singularities in Feynman--Wheeler and fakeon prescriptions}
\label{subsec:FW_fakeon_second_type}

{We now recall how the second type of singularities arises in the Feynman--Wheeler construction
and how the same mechanism is different in the dual-IHO theory. This point is more easily
seen in a scalar two-point bubble. Consider two timelike denominators,
\begin{equation}
D_1(k)=k^2+m_1^2,
\qquad
D_2(P-k)=(P-k)^2+m_2^2,
\end{equation}
with our signature $k^2=-k_0^2+\mathbf{k}^2$. The ordinary Feynman bubble is
\begin{equation}
I_{\rm F}(P)
=
\int {d^4 k\over (2\pi)^4}\,
{1\over D_1(k)-i\epsilon}\,
{1\over D_2(P-k)-i\epsilon}.
\end{equation}
Both poles carry the same boundary prescription. After Feynman parametrization,
\begin{equation}
{1\over (D_1-i\epsilon)(D_2-i\epsilon)}
=
\int_0^1 dx\,
{1\over \left[xD_1+(1-x)D_2-i\epsilon\right]^2}.
\end{equation}
The sign of the infinitesimal imaginary part is independent of $x$. The Wick rotation and
the ultraviolet subtraction are therefore the usual ones. The divergent part comes from the
large-$k$ expansion and is local. The Feynman--Wheeler prescription is different. One replaces one of the internal lines by
a Cauchy principal value,
\begin{equation}
{\rm PV}\left({1\over D_2}\right)
=
{1\over2}
\lim_{\epsilon\to0^+}
\left[
{1\over D_2-i\epsilon}
+
{1\over D_2+i\epsilon}
\right].
\end{equation}
Thus, the Feynman--Wheeler bubble is the half-sum
\begin{equation}
I_{\rm FW}(P)
=
{1\over2}I_{--}(P)
+
{1\over2}I_{-+}(P),
\end{equation}
where
\begin{equation}
I_{--}(P)
=
\int {d^4 k\over (2\pi)^4}\,
{1\over D_1-i\epsilon}\,
{1\over D_2-i\epsilon},
\end{equation}
and
\begin{equation}
I_{-+}(P)
=
\int {d^4 k\over (2\pi)^4}\,
{1\over D_1-i\epsilon}\,
{1\over D_2+i\epsilon}.
\end{equation}
The first term is an ordinary Feynman integral. The second term contains opposite
boundary values in the same loop.
For $I_{-+}$, the Feynman parametrization gives
\begin{equation}
{1\over (D_1-i\epsilon)(D_2+i\epsilon)}
=
\int_0^1 dx\,
{1\over
\left[
xD_1+(1-x)D_2+i\epsilon(1-2x)
\right]^2}.
\end{equation}
The coefficient of $i\epsilon$ vanishes at $x={1\over2}$ and signals a pathology. The combined denominator does not carry a uniform boundary prescription over the Feynman parameter domain. Equivalently, the two sets of poles can approach the loop-energy contour from opposite sides. This is the
mechanism by which a pinch can occur at the boundary of loop-momentum space. To see that this is a second type of singularity, take the projective large-momentum limit
\begin{equation}
k^\mu=\Lambda n^\mu,
\qquad
\Lambda\to\infty ,
\end{equation}
then
\begin{equation}
D_1(k)
=
\Lambda^2 n^2+m_1^2,
\end{equation}
and
\begin{equation}
D_2(P-k)
=
\Lambda^2 n^2-2\Lambda P\cdot n+P^2+m_2^2 .
\end{equation}
At the leading order, the masses drop out. A pinch at infinity requires the two leading
surfaces to degenerate. This gives the projective conditions
\begin{equation}
n^2=0,
\qquad
P\cdot n=0 .
\end{equation}
For a nonzero real null direction $n^\mu$, these conditions force the external momentum
to lie on the special locus
\begin{equation}
P^2=0 .
\end{equation}
Thus, the singularity is not an ordinary massive threshold. It is a singularity at null
infinity in loop-momentum space. This is what is meant by a second type of Landau
singularity.
The resulting divergent term \eqref{eq:nonlocal} is nonlocal because the singular locus is $P^2=0$, rather
than a polynomial expansion around large $k$. 
The factor $1/P^2$ in \eqref{eq:nonlocal} is the signature of the second type singularity. It is not a local
counterterm. A local ultraviolet counterterm would be a polynomial in the external
momentum, such as
\begin{equation}
1,\qquad P^2,\qquad P^4 .
\label{PPowers}
\end{equation}
The Feynman--Wheeler prescription therefore, fails because the coexistence of opposite
$i\epsilon$ prescriptions at timelike poles can turn a UV-boundary pinch into a nonlocal
divergence.
The fakeon prescription was introduced precisely to avoid this problem. In the fakeon
construction, one does not define the loop amplitude by inserting the naive Feynman--Wheeler
principal value directly inside Minkowski loop integrals. Instead, one starts from the
Euclidean amplitude performs the usual local renormalization there, and only then
continues back to Minkowski space using the average continuation across fakeon thresholds.
Schematically,
\begin{equation}
\mathcal{A}_{\rm fake}(P)
=
{1\over2}
\left[
\mathcal{A}_{+}(P)
+
\mathcal{A}_{-}(P)
\right]_{\rm after\ Euclidean\ renormalization}.
\end{equation}
The order of operations is essential. In the Feynman--Wheeler case, the average is inserted
inside the loop integral before renormalization,
\begin{equation}
\int d^4k\,
{1\over D_1-i\epsilon}\,
{1\over2}
\left[
{1\over D_2-i\epsilon}
+
{1\over D_2+i\epsilon}
\right].
\end{equation}
In the fakeon case, the ultraviolet divergence is first subtracted in the Euclidean theory,
where the large-momentum expansion is local, and only afterwards is the average
continuation applied:
\begin{equation}
\left[
\int d^4k_E\,
{1\over D_{1,E}}\,
{1\over D_{2,E}}
\right]_{\rm ren}
\quad
\longrightarrow
\quad
{\rm average\ continuation}.
\end{equation}
Thus, the fakeon prescription does not generate the Feynman--Wheeler nonlocal UV
counterterm. It removes the absorptive part associated with the fakeon threshold, but the
UV counterterms remain those obtained from the local Euclidean expansion.}

{This comparison is important for the dual-IHO construction. 
The three cases should therefore be kept distinct. The Feynman--Wheeler prescription inserts opposite timelike boundary values directly in Minkowski loop integrals and can generate second-type nonlocal divergences. The fakeon prescription avoids this by renormalizing the Euclidean amplitude first and applying average continuation only after
local subtraction. The dual-IHO construction is different from both: its pole is spacelike, its on-shell spectral term is absent by the Wightman/KL condition, and its
ultraviolet subtraction is performed on the full local four-derivative propagator.}

\begin{table}[b]
\centering
\renewcommand{\arraystretch}{1.35}
\begin{tabular}{lccc}
\toprule
Bubble & Pole types & $\alpha_2/\alpha_1$ & Threshold\\
\midrule
Dual-IHO [this work] & both spacelike & $1\in\mathbb R^+$
  & $P^2_*=4\mu_2^2>0$ (spacelike)\\
Mixed [this work] & one each & $i\mu_2/m\in i\mathbb R$
  & complex (no physical pinch)\\
FW~\cite{Anselmi2020} & both timelike & $m_2/m_1\in\mathbb R^+$
  & $P^2=-(m_1{\pm}m_2)^2<0$, nonlocal div.\\
\bottomrule
\end{tabular}
\caption{Landau-equation comparison. Nonlocal divergences arise only when both poles are
timelike (FW case) and $\alpha_2/\alpha_1$ is real positive, giving a physical threshold.
Spacelike poles drive the ratio complex, eliminating any physical pinch.}
\label{tab:landau}
\end{table}

\subsection{{Locality of counterterms and absence of second type nonlocal divergences}}

\label{sec:dualIHOdiver}

{The ultraviolet behavior of the dual-IHO spin--2 sector must be read from the full quadratic gravity propagator, not from an artificial decomposition of the principal value
distribution into separate Feynman and anti-Feynman boundary values. The dual-IHO Green distribution is
\begin{equation}
\Delta_{\rm dIHO}(k)
=
i\,{\rm PV}\left({1\over k^2-\mu_2^2}\right) \underset{\textrm{phys.}}{\neq} \frac{1}{2}\Bigg[\frac{i}{k^2-\mu_2^2+i\epsilon}+\frac{i}{k^2-\mu_2^2-i\epsilon}\Bigg]
\end{equation}
where the principal value form follows from the absence of physical spectral weight on the spacelike shell $k^2=\mu_2^2>0$. As discussed in Sec.~\ref{sec:KL} let us recall that the principal value has a more fundamental definition that is not associated with (anti-)Feynman prescriptions that require particle interpretations and the existence of a vacuum \cite{GelfandShilov1964,BogoliubovShirkov1980}. Therefore, the PV of dual-IHO is not physically equal to the average of Feynman and anti-Feynman type contributions for the present case (See Appendix.~\ref{subsec:not_FW_prescription} for more details). At large momentum, the principal value distribution has the same asymptotic expansion as
the corresponding rational function:
\begin{equation}
{\rm PV}\left({1\over k^2-\mu_2^2}\right)
=
{1\over k^2}
\left(
1+{\mu_2^2\over k^2}+O(k^{-4})
\right),
\qquad |k^2|\to\infty .
\label{eq:1k4}
\end{equation}
The difference between the Feynman and principal value boundary values is
\begin{equation}
{1\over k^2-\mu_2^2-i\epsilon}
-
{\rm PV}\left({1\over k^2-\mu_2^2}\right)
=
i\pi\delta(k^2-\mu_2^2).
\label{eq:FPV}
\end{equation}
This difference is supported on the finite spacelike shell $k^2=\mu_2^2$. It is the on-shell delta contribution, namely the cut-producing part, which is absent in the dual-IHO
sector because the physical KL spectral weight on this shell vanishes. It is not an ultraviolet contribution.
The full spin--2 propagator has the schematic form
\begin{equation}
D^{(2)}_{\mu\nu\rho\sigma}(k)
=
-{i\over M_{\rm P}^2}
P^{(2)}_{\mu\nu\rho\sigma}
\left[
{1\over k^2-i\epsilon}
-
{\rm PV}\left({1\over k^2-\mu_2^2}\right)
\right],
\end{equation}
which we recall that at large momentum $\sim
-{\mu_2^2\over k^4}$,
and the full spin--2 propagator retains the $1/k^4$ ultraviolet falloff responsible for
Stelle power-counting renormalizability.
The remaining question is whether a second type of singularity at the boundary of loop-momentum space can generate a nonlocal ultraviolet divergence. A second type
of singularity is not an ordinary particle production threshold. It corresponds to a possible
pinch of the loop integration at $k^\mu\to\infty$. Such a singularity, if it produced an
uncancelled ultraviolet divergence, would appear as a nonlocal term such as \eqref{eq:nonlocal}.
This would not be a local counterterm.
In the dual-IHO theory, this term is not generated. The reason is that the ultraviolet
subtraction is performed on the full local four-derivative propagator. In the ultraviolet
region $k^\mu\gg P^\mu$, the integrand is expanded in powers of the external momentum.
For example,
\begin{equation}
{1\over (k-P)^2-\mu^2}
=
{1\over k^2-\mu^2}
+
{2k\cdot P-P^2\over (k^2-\mu^2)^2}
+
{(2k\cdot P-P^2)^2\over (k^2-\mu^2)^3}
+\cdots .
\end{equation}
The same asymptotic expansion holds distributionally for the principal value inverse. Each
term in the large-$k$ expansion is polynomial in $P^\mu$. Hence, the divergent part of the
loop integral is polynomial in the external momentum. It can generate local terms
proportional to \eqref{PPowers}, but it cannot generate a non-polynomial ultraviolet counterterm proportional to \eqref{eq:nonlocal}.
This is the point at which the dual-IHO construction differs from a Feynman--Wheeler or
fakeon average of timelike propagators. In the latter case, one decomposes a timelike pole
into opposite $i\epsilon$ boundary values, and the coexistence of these boundary values can
pinch the loop contour at infinity, producing a second type of nonlocal term. In the present
case, the internal line is not such an average as we also argued in Appendix.~\ref{subsec:not_FW_prescription}. It is the principal value of the Green distribution of
a spacelike dual-IHO pole with no physical spectral delta function. The UV expansion is
therefore the local expansion of the full four-derivative propagator, and its divergent part
is polynomial.
Consequently, the dual-IHO spin--2 sector has no physical Cutkosky cut and no nonlocal
ultraviolet counterterm of the form \eqref{eq:nonlocal}.
The allowed counterterms are the local generally covariant terms with at most four
derivatives, equivalently
\begin{equation}
\int d^4x\sqrt{-g},
\qquad
\int d^4x\sqrt{-g}\,R,
\qquad
\int d^4x\sqrt{-g}\,R^2,
\qquad
\int d^4x\sqrt{-g}\,R_{\mu\nu}R^{\mu\nu},
\end{equation}
or a basis involving $C_{\mu\nu\rho\sigma}C^{\mu\nu\rho\sigma}$ and the Gauss--Bonnet
density. The locality of counterterms and the $1/k^4$ ultraviolet falloff are therefore
preserved.}

\subsection{Comparison with existing approaches}
\label{sec:compare}

For $\beta<0$ (ghost), the timelike pole carries a wrong-sign residue with $\rho\neq0$: the Feynman propagator is admissible, but its imaginary part violates the optical theorem. The fakeon~\cite{Anselmi2018,Melis:2022tqz} removes the ghost from asymptotic states by average continuation from Euclidean space, abandoning analyticity above threshold and also leading to microcausality violation. The Lee--Wick prescription~\cite{LeeWick,Donoghue2019} modifies the $k^0$ contour or interprets the ghost as an unstable resonance, and the PT-symmetric quantization~\cite{Bender2008} redefines the inner product. All the mentioned prescriptions start from a timelike pole and ask how to virtualize it. For $\beta>0$ (dual-IHO), the pole is spacelike and consequently $\rho=0$. There is no Feynman propagator to start from, no ghost to remove, no contour to deform. The PV propagator \eqref{eq:PV} is the only propagator the theory admits (due to the KL representation failure), and its cut vanishes, not by construction but by necessity.

\section{Conclusions}
\label{sec:conclusions}

Quadratic gravity is the unique perturbatively renormalizable local theory of quantum gravity in four spacetime dimensions, but its extra spin--2 sector has long been regarded as the source of non-unitarity. The result of this paper is that this conclusion is not inevitable. For a positive Weyl-squared coefficient, the additional spin--2 mode is not a ghost associated with a negative definite Hamiltonian but a dual inverted harmonic oscillator (IHO) associated with an indefinite Hamiltonian (equivalent to IHO). It is worth noting that IHO is a universal physical system that appears across different areas of physics \cite{Subramanyan:2020fmx,Sundaram:2024ici,Gaztanaga2026,Kumar:2026ggw}. 
Starting with the dual inverted harmonic oscillator (IHO) condition ($\beta>0$), the additional spin-2 sector is characterized by a non-normalizable ground state and the presence of a spacelike pole. These two facts force the spectral density of the dual IHO spin-2 field $\rho$ to vanish entirely, which in turn reduces the propagator $\Delta$ to its PV part. Ultimately, this implies that the theory lacks the necessary branch cuts associated with physical particle production which indicate the dual IHO spin-2 cannot have any particle interpretation, no vacuum, not an asymptotic state but dictates the virtual physical processes. The main results obtained here can be summarized as:

\begin{itemize}
\item For a dual IHO spin--2 field the Wightman spectrum condition requires all physical states to have real non-negative mass. A spacelike pole $k^2=\mu_2^2>0$ would require $s=-\mu_2^2<0$ in the KL integral, which is impossible. Independently, no normalizable vacuum exists. Both conditions, independently, force the spectral density to vanish $\rho=0$.

\item The principal value propagator is unique. Eq.~\eqref{eq:PV} is the only KL consistent propagator for the dual-IHO. The identity $\PV=\frac12(F+\bar F)$ is algebraically valid but is empty of physical meaning, since neither the Feynman nor the anti-Feynman propagators are allowed for $\rho=0$.

\item The dual-IHO bubble has threshold $P^2_*=4\mu_2^2>0$ (spacelike), kinematically inaccessible for physical processes (Eq.~\eqref{eq:BP}).

\item The mixed bubble (one timelike, one spacelike pole) gives $\alpha_2/\alpha_1=i\mu_2/m$ (which is purely imaginary, see Eq.~\eqref{eq:complex_L}), therefore there is no real kinematic pinch and hence no nonlocal divergence. This is the decisive structural difference from the Feynman--Wheeler case~\cite{Anselmi2020} (Table~\ref{tab:landau}).

\item The Feynman--PV propagator difference \eqref{eq:FPV} has no UV support, locality of counterterms and the $1/k^4$ scaling \eqref{eq:1k4} remain intact. Therefore, renormalizability preserved.
\end{itemize}

Unitarity and renormalizability coexist in quadratic gravity for $\beta>0$ without any prescription beyond the structure of the theory. The DQFT quantization framework underlying this construction, including direct-sum Hilbert spaces, geometric superselection sectors, and the resolution of the two-arrows-of-time problem of the IHO identified by Berry and Keating, is developed in Refs.~\cite{KumarMarto2026,Kumar2025dS,Kumar2024BH,Kumar2024Rindler,Gaztanaga2026}, together with cosmological implications for inflationary gravitational waves and CMB parity asymmetry. Thus, we have a unique unitary quantum gravity in 4D that is predictive and testable with future observations. 

\acknowledgments

K.S.K.\ acknowledges support from the Royal Society Newton International Fellowship.
Research funded by FCT grant UIDB/MAT/00212/2025 and COST action 23130. We thank David Wands for the positive encouragement and suggestion to write this short paper. We also thank Marco Piva, Krzysztof Jodlowski for the useful comments on our companion paper \cite{KumarMarto2026} related to the aspects of PV propagator. We also thank Luca  Buoninfante for useful discussions over the years. 

\appendix

\section{Why the dual-IHO propagator is not a Feynman--Wheeler prescription}
\label{subsec:not_FW_prescription}

{We now make precise why the principal value propagator of the dual-IHO spin--2 sector is
not a Feynman--Wheeler prescription. The distinction is important. The Feynman--Wheeler
construction starts from timelike particle poles and averages opposite boundary values.
The dual-IHO construction starts instead from the Green distribution of a local quadratic
operator whose additional spin--2 pole is spacelike.
Suppressing spin--2 projectors and overall normalization factors, the dual-IHO Green
function is defined in position space by
\begin{equation}
(\Box+\mu_2^2)\widetilde{\Delta}_{\rm dIHO}(x-y)
=
-i\,\delta^{(4)}(x-y).
\end{equation}
We use the Fourier convention
\begin{equation}
\widetilde{\Delta}_{\rm dIHO}(x-y)
=
\int {d^4k\over (2\pi)^4}\,
e^{ik\cdot(x-y)}
\Delta_{\rm dIHO}(k).
\end{equation}
With signature
\begin{equation}
k^2=-k_0^2+\mathbf{k}^2,
\end{equation}
one has
\begin{equation}
\Box\rightarrow -k^2 .
\end{equation}
Therefore the Green-function equation becomes
\begin{equation}
(k^2-\mu_2^2)\Delta_{\rm dIHO}(k)=i .
\end{equation}
Equivalently, writing
\begin{equation}
X\equiv k^2-\mu_2^2,
\end{equation}
one has the distributional equation
\begin{equation}
X\,\Delta(X)=i .
\end{equation}
This statement requires no reference to a vacuum, a Fock space, or a
Källén--Lehmann representation. It is simply the statement that
$\Delta_{\rm dIHO}(k)$ is a right inverse of the local kinetic operator.
A particular solution is
\begin{equation}
\Delta_{\rm part}(X)
=
i\,{\rm PV}\left({1\over X}\right),
\end{equation}
because the Cauchy principal value satisfies
\begin{equation}
X\,{\rm PV}\left({1\over X}\right)=1
\end{equation}
as a distribution. The algebraic inverse, however, is not yet unique. If
$\Delta_1$ and $\Delta_2$ are two solutions, their difference
$T=\Delta_1-\Delta_2$ obeys
\begin{equation}
X\,T(X)=0 .
\end{equation}
Thus $T$ must be supported on the hypersurface $X=0$. Among Lorentz invariant
distributions supported on this hypersurface, the only one annihilated by multiplication by
$X$ is proportional to $\delta(X)$. Derivatives such as $\delta'(X)$ are not annihilated,
since multiplication by $X$ lowers the derivative order. Hence, the most general
Lorentz invariant Green distribution is
\begin{equation}
\Delta(X)
=
i\,{\rm PV}\left({1\over X}\right)
+
C\,\delta(X).
\end{equation}
Restoring $X=k^2-\mu_2^2$, this becomes
\begin{equation}
\Delta_{\rm dIHO}(k)
=
i\,{\rm PV}\left({1\over k^2-\mu_2^2}\right)
+
C\,\delta(k^2-\mu_2^2).
\end{equation}
The constant $C$ is not fixed by the local Green function equation. It is fixed by the
spectral condition. The homogeneous term is supported on
\begin{equation}
k^2=\mu_2^2>0 .
\end{equation}
This is a spacelike shell. In a Källén--Lehmann representation the physical spectral
variable is
\begin{equation}
s=-k^2,
\end{equation}
with physical support restricted to
\begin{equation}
s\geq0 .
\end{equation}
The shell $k^2=\mu_2^2$ would correspond to
\begin{equation}
s=-\mu_2^2<0,
\end{equation}
which lies outside the Wightman spectral domain. Therefore, this shell cannot carry
physical on-shell spectral weight. The homogeneous spectral coefficient must vanish,
\begin{equation}
C=0 .
\end{equation}
Thus, the dual-IHO momentum-space Green distribution is
\begin{equation}
\Delta_{\rm dIHO}(k)
=
i\,{\rm PV}\left({1\over k^2-\mu_2^2}\right).
\end{equation}
This derivation is independent of any Feynman--Wheeler averaging rule applied in the context of theory with ghost $\beta<0$ (i.e., the massive spin--2 with negative definite Hamiltonian). To see the difference, recall the distributional identities \begin{equation} \lim_{\epsilon\to0^+}{i\over X+i\epsilon} = i\,{\rm PV}\left({1\over X}\right) +\pi\delta(X), \qquad \lim_{\epsilon\to0^+}{i\over X-i\epsilon} = i\,{\rm PV}\left({1\over X}\right) -\pi\delta(X). \end{equation} Therefore \begin{equation} i\,{\rm PV}\left({1\over X}\right) = {1\over2} \lim_{\epsilon\to0^+} \left[ {i\over X+i\epsilon} + {i\over X-i\epsilon} \right]. \end{equation} This equality is an identity between distributions. It is not the physical definition of the dual-IHO propagator. The two individual boundary values contain nonzero homogeneous terms proportional to $\delta(X)$. In the present case these terms live on the spacelike shell $k^2=\mu_2^2>0$, precisely where the Wightman/Källén--Lehmann spectral condition forbids physical on-shell support. Thus the principal-value form is not obtained by averaging two allowed particle propagators. It is obtained by solving the local Green-function equation and then imposing the absence of physical spectral weight on the spacelike shell. This is the mathematical difference from a Feynman--Wheeler construction. In a Feynman--Wheeler or fakeon treatment, one begins with timelike particle poles carrying opposite $i\epsilon$ boundary values and averages them. In the dual-IHO case, the pole is spacelike from the outset, and the possible on-shell spectral term is removed by the Wightman spectral condition. The notation \begin{equation} i\,{\rm PV}\left({1\over k^2-\mu_2^2}\right) \end{equation} therefore denotes the Green distribution selected by the local kinetic operator and the spectral condition, not a prescription imposed on a timelike ghost pole as it is usually done in the case of $\beta<0$ \cite{Anselmi2018}.}

\bibliographystyle{utphys}
\bibliography{QQG.bib}

@article{Stelle1977,
  author       = {Stelle, K. S.},
  title        = {Renormalization of Higher-Derivative Quantum Gravity},
  journal      = {Phys. Rev. D},
  volume       = {16},
  pages        = {953--969},
  year         = {1977},
  doi          = {10.1103/PhysRevD.16.953}
}

@misc{KumarMarto2026,
  author       = {Kumar, K. S. and Marto, J.},
  title        = {Quantum (quadratic) gravity: replacing the massive tensor ghost with an inverted harmonic oscillator-like instability},
  year         = {2026},
  eprint       = {2603.07150},
  archivePrefix= {arXiv},
  primaryClass = {hep-th},
  url          = {https://arxiv.org/abs/2603.07150}
}

@article{Buoninfante2025,
  author       = {Buoninfante, L.},
  title        = {Strict renormalizability as a paradigm for fundamental physics},
  journal      = {JHEP},
  volume       = {07},
  pages        = {175},
  year         = {2025},
  eprint       = {2504.05900},
  archivePrefix= {arXiv},
  primaryClass = {hep-th},
  url          = {https://arxiv.org/abs/2504.05900}
}

@article{Salvio2018,
  author       = {Salvio, A.},
  title        = {Quadratic Gravity},
  journal      = {Front. Phys.},
  volume       = {6},
  pages        = {77},
  year         = {2018},
  eprint       = {1804.09944},
  archivePrefix= {arXiv},
  primaryClass = {hep-th},
  url          = {https://arxiv.org/abs/1804.09944}
}

@article{Anselmi2018,
  author       = {Anselmi, D. and Piva, M.},
  title        = {Quantum Gravity, Fakeons and Microcausality},
  journal      = {JHEP},
  volume       = {11},
  pages        = {021},
  year         = {2018},
  eprint       = {1806.03605},
  archivePrefix= {arXiv},
  primaryClass = {hep-th},
  url          = {https://arxiv.org/abs/1806.03605}
}

@article{Anselmi2020,
  author       = {Anselmi, D.},
  title        = {The quest for purely virtual quanta: fakeons versus Feynman--Wheeler particles},
  journal      = {JHEP},
  volume       = {03},
  pages        = {142},
  year         = {2021},
  eprint       = {2001.01942},
  archivePrefix= {arXiv},
  primaryClass = {hep-th},
  url          = {https://arxiv.org/abs/2001.01942}
}

@article{LeeWick,
  author       = {Lee, T. D. and Wick, G. C.},
  title        = {Negative Metric and the Unitarity of the S-Matrix},
  journal      = {Nucl. Phys. B},
  volume       = {9},
  pages        = {209--243},
  year         = {1969},
  doi          = {10.1016/0550-3213(69)90098-4}
}

@article{Donoghue2019,
  author       = {Donoghue, J. F. and Menezes, G.},
  title        = {Arrow of Causality and Quantum Gravity},
  journal      = {Phys. Rev. Lett.},
  volume       = {123},
  pages        = {171601},
  year         = {2019},
  eprint       = {1908.04170},
  archivePrefix= {arXiv},
  primaryClass = {hep-th},
  url          = {https://arxiv.org/abs/1908.04170}
}

@article{Bender2008,
  author       = {Bender, C. M. and Mannheim, P. D.},
  title        = {No-ghost theorem for the fourth-order derivative Pais--Uhlenbeck oscillator model},
  journal      = {Phys. Rev. Lett.},
  volume       = {100},
  pages        = {110402},
  year         = {2008},
  eprint       = {0706.0207},
  archivePrefix= {arXiv},
  primaryClass = {hep-th},
  url          = {https://arxiv.org/abs/0706.0207}
}

@incollection{BerryKeating1999,
  author       = {Berry, M. V. and Keating, J. P.},
  title        = {{$H=xp$ and the Riemann Zeros}},
  booktitle    = {Supersymmetry and Trace Formulae},
  publisher    = {Springer},
  pages        = {355--367},
  year         = {1999},
  doi          = {10.1007/978-1-4615-4875-1_19}
}

@article{Kallen1952,
  author       = {K{\"a}ll{\'e}n, G.},
  title        = {On the definition of the Renormalization Constants in Quantum Electrodynamics},
  journal      = {Helv. Phys. Acta},
  volume       = {25},
  pages        = {417--434},
  year         = {1952}
}

@article{Lehmann1954,
  author       = {Lehmann, H.},
  title        = {On the Properties of propagation functions and renormalization constants of quantized fields},
  journal      = {Nuovo Cim.},
  volume       = {11},
  pages        = {342--357},
  year         = {1954},
  doi          = {10.1007/BF02783624}
}

@book{Streater1964,
  author       = {Streater, R. F. and Wightman, A. S.},
  title        = {PCT, Spin and Statistics, and All That},
  publisher    = {Princeton University Press},
  year         = {1964}
}

@book{Haag1996,
  author       = {Haag, R.},
  title        = {Local Quantum Physics: Fields, Particles, Algebras},
  publisher    = {Springer},
  edition      = {2},
  year         = {1996},
  doi          = {10.1007/978-3-642-61458-3}
}

@article{Jacobson1988,
  author       = {Jacobson, T. and Tsamis, N. C. and Woodard, R. P.},
  title        = {Tachyons and Perturbative Unitarity},
  journal      = {Phys. Rev. D},
  volume       = {38},
  pages        = {1823--1836},
  year         = {1988},
  doi          = {10.1103/PhysRevD.38.1823}
}

@article{Kumar2025dS,
  author       = {Kumar, K. S. and Marto, J.},
  title        = {Towards a Unitary Formulation of Quantum Field Theory in Curved Spacetime: The Case of de Sitter Spacetime},
  journal      = {Symmetry},
  volume       = {17},
  pages        = {29},
  year         = {2025},
  eprint       = {2305.06046},
  archivePrefix= {arXiv},
  primaryClass = {hep-th},
  url          = {https://arxiv.org/abs/2305.06046}
}

@article{Kumar2024BH,
  author       = {Kumar, K. S. and Marto, J.},
  title        = {Towards a Unitary Formulation of Quantum Field Theory in Curved Space-Time: The Case of the Schwarzschild Black Hole},
  journal      = {PTEP},
  volume       = {2024},
  pages        = {123E01},
  year         = {2024},
  eprint       = {2307.10345},
  archivePrefix= {arXiv},
  primaryClass = {hep-th},
  url          = {https://arxiv.org/abs/2307.10345}
}

@article{Kumar2024Rindler,
  author       = {Kumar, K. S. and Marto, J.},
  title        = {Revisiting Quantum Field Theory in Rindler Spacetime with Superselection Rules},
  journal      = {Universe},
  volume       = {10},
  pages        = {320},
  year         = {2024},
  eprint       = {2405.20995},
  archivePrefix= {arXiv},
  primaryClass = {hep-th},
  url          = {https://arxiv.org/abs/2405.20995}
}

@article{Kumar2024Hawking,
  author       = {Kumar, K. S. and Marto, J.},
  title        = {Hawking radiation with pure states},
  journal      = {Gen. Rel. Grav.},
  volume       = {56},
  pages        = {143},
  year         = {2024},
  eprint       = {2407.18652},
  archivePrefix= {arXiv},
  primaryClass = {hep-th},
  url          = {https://arxiv.org/abs/2407.18652}
}

@article{Gaztanaga2026,
  author       = {Gazta{\~n}aga, E. and Kumar, K. S. and Marto, J.},
  title        = {A new understanding of Einstein--Rosen bridges},
  journal      = {Class. Quant. Grav.},
  volume       = {43},
  pages        = {015023},
  year         = {2026},
  eprint       = {2512.20691},
  archivePrefix= {arXiv},
  primaryClass = {gr-qc},
  url          = {https://arxiv.org/abs/2512.20691}
}

@book{WeinbergQFT1,
  author       = {Weinberg, Steven},
  title        = {The Quantum Theory of Fields. Volume I: Foundations},
  publisher    = {Cambridge University Press},
  year         = {1995}
}

@article{Jodlowski:2026zaw,
    author = "Jod{\l}owski, Krzysztof",
    title = "{Is a covariant virtual tachyon viable?}",
    eprint = "2602.20474",
    archivePrefix = "arXiv",
    primaryClass = "hep-ph",
    reportNumber = "CTPU-PTC-26-06",
    doi = "10.1103/spbd-sd3s",
    journal = "Phys. Rev. D",
    volume = "113",
    number = "6",
    pages = "065016",
    year = "2026"
}

@article{Subramanyan:2020fmx,
    author = "Subramanyan, Varsha and Hegde, Suraj S. and Vishveshwara, Smitha and Bradlyn, Barry",
    title = "{Physics of the Inverted Harmonic Oscillator: From the lowest Landau level to event horizons}",
    eprint = "2012.09875",
    archivePrefix = "arXiv",
    primaryClass = "cond-mat.mes-hall",
    doi = "10.1016/j.aop.2021.168470",
    journal = "Annals Phys.",
    volume = "435",
    pages = "168470",
    year = "2021"
}

@article{Sundaram:2024ici,
    author = "Sundaram, Sriram and Burgess, C. P. and O'Dell, D. H. J.",
    title = "{Duality between the quantum inverted harmonic oscillator and inverse square potentials}",
    eprint = "2402.13909",
    archivePrefix = "arXiv",
    primaryClass = "quant-ph",
    doi = "10.1088/1367-2630/ad3a91",
    journal = "New J. Phys.",
    volume = "26",
    number = "5",
    pages = "053023",
    year = "2024"
}

@article{Melis:2022tqz,
    author = "Melis, Aurora and Piva, Marco",
    title = "{One-loop integrals for purely virtual particles}",
    eprint = "2209.05547",
    archivePrefix = "arXiv",
    primaryClass = "hep-ph",
    doi = "10.1103/PhysRevD.108.096021",
    journal = "Phys. Rev. D",
    volume = "108",
    number = "9",
    pages = "096021",
    year = "2023"
}

@article{Jacobson:1987ap,
    author = "Jacobson, T. and Tsamis, N. C. and Woodard, R. P.",
    title = "{Tachyons and Perturbative Unitarity}",
    reportNumber = "BRX-TH-231, BROWN-HET-642",
    doi = "10.1103/PhysRevD.38.1823",
    journal = "Phys. Rev. D",
    volume = "38",
    pages = "1823",
    year = "1988"
}

@article{Kumar:2026ggw,
    author = "Kumar, K. Sravan",
    title = "{The Saddle Point of Everything:~The inverted harmonic oscillator, the universal physics of unstable equilibria, and the unique renormalizable theory of quantum gravity}",
    eprint = "2605.30386",
    archivePrefix = "arXiv",
    primaryClass = "physics.gen-ph",
    month = "5",
    year = "2026"
}

@book{GelfandShilov1964,
  author    = {Gel'fand, I. M. and Shilov, G. E.},
  title     = {Generalized Functions. Volume 1: Properties and Operations},
  publisher = {Academic Press},
  address   = {New York},
  year      = {1964}
}

@book{BogoliubovShirkov1980,
  author    = {N. N. Bogoliubov and D. V. Shirkov},
  title     = {Introduction to the Theory of Quantized Fields},
  edition   = {3},
  publisher = {John Wiley \& Sons},
  address   = {New York},
  year      = {1980},
  isbn      = {9780471040794}
}

\end{document}